\documentclass[%
reprint,
showpacs,
bibnotes,
amsmath,
amssymb,
aps,
prb,
floatfix
]{revtex4-1}

\usepackage{graphicx}
\usepackage{dcolumn}
\usepackage{bm}
\usepackage{hyperref}
\usepackage{multirow}
\usepackage{array}
\usepackage{booktabs}
\usepackage{ctable}
\usepackage{upgreek}
\usepackage{epsfig}
\usepackage{mathrsfs}
\usepackage{amssymb}
\usepackage{amsbsy}
\usepackage{color}
\usepackage{cancel}
\usepackage{marginnote}
\newcommand{\bcen}{\begin{center}}
\newcommand{\ecen}{\end{center}}
\newcommand{\btab}{\begin{tabular}}
\newcommand{\etab}{\end{tabular}}
\newcommand{\bdes}{\begin{description}}
\newcommand{\edes}{\end{description}}

\newcommand{\beq}{\begin{equation}}
\newcommand{\eeq}{\end{equation}}
\newcommand{\bea}{\begin{eqnarray}}
\newcommand{\eea}{\end{eqnarray}}

\newcommand{\half}{\frac{1}{2}}
\newcommand{\bary}{\begin{array}}
\newcommand{\eary}{\end{array}}
\newcommand{\benum}{\begin{enumerate}}
\newcommand{\eenum}{\end{enumerate}}
\newcommand{\bitem}{\begin{itemize}}
\newcommand{\eitem}{\end{itemize}}

\newcommand{\btau}{\mbox{\boldmath $ \tau $}}
\newcommand{\blam}{{\boldsymbol{\lambda}}}

\newcommand{\bOne}{{\boldsymbol 1}}
\newcommand{\be} { \mbox{\boldmath $e$}}

\newcommand{\bk} { \bm{k} }

\newcommand{\bp} { \bm{p} }
\newcommand{\bq} { \bm{q} }
\newcommand{\br} { \boldsymbol{r}}

\newcommand{\bzero} { {\boldsymbol{0}}}

\newcommand{\dou}{\partial}

\newcommand{\D}[1]{\mbox{d}{#1}}

\newcommand{\ket}[1]{| #1 \rangle}

\newcommand{\eqn}[1] {eqn.~(\ref{#1})}

\newcommand{\fig}[1]{fig.~\ref{#1}}
\newcommand{\Fig}[1]{Fig.~\ref{#1}}

\makeatletter

\newcommand{\Rmnum}[1]{\expandafter\@slowromancap\romannumeral #1@}
\makeatother

\newcommand{\myfigwidth}{0.4\paperwidth}

\newcommand{\asc}{a_{sc}}
\newcommand{\as}{a_{s}}

\newcommand{\Ef}{E_F}
\newcommand{\kf}{k_F}

\newcommand{\mylabel}[1]{\label{#1}} 

\newcommand{\myonlinecite}[1]{[\onlinecite{#1}]}

\usepackage{lineno}

\begin{document}


\title{Rashbons: Properties and their significance}
\date{August 24, 2011}
\author{Jayantha P. Vyasanakere}
\email{jayantha@physics.iisc.ernet.in}
\author{Vijay B. Shenoy}
\email{shenoy@physics.iisc.ernet.in}
\affiliation{Centre for Condensed Matter Theory, Department of Physics, Indian Institute of Science, Bangalore 560 012, India}



\begin{abstract}

In presence of a synthetic non-Abelian gauge field that induces a
Rashba like spin-orbit interaction, a collection of weakly interacting
fermions undergoes a crossover from a BCS ground state to a BEC ground
state when the strength of the gauge field is increased [Phys.~Rev.~B
  {\bf 84}, 014512 (2011)]. The BEC that is obtained at large gauge
coupling strengths is a condensate of tightly bound bosonic
fermion-pairs whose properties are solely determined by the Rashba
gauge field -- hence called rashbons. In this paper, we conduct a
systematic study of the properties of rashbons and their dispersion.
This study reveals a new qualitative aspect of the problem of
interacting fermions in non-Abelian gauge fields, i.~e, that the
rashbon state induced by the gauge field for small centre of mass
momenta of the fermions ceases to exist when this momentum exceeds a
critical value which is of the order of the gauge coupling
strength. The study allows us to estimate the transition temperature
of the rashbon BEC, and suggests a route to enhance the exponentially
small transition temperature of the system with a fixed weak
attraction to the order of the Fermi temperature by tuning the
strength of the non-Abelian gauge field.  The nature of the rashbon
dispersion, and in particular the absence of the rashbon states at large
momenta, suggests a regime of parameter space where the normal state
of the system will be a dynamical mixture of uncondensed rashbons and
unpaired helical fermions. Such a state should show many novel
features including pseudogap physics.

\end{abstract}


\maketitle

\section{Introduction}
\mylabel{sec:Intro}

Cold atoms are a promising platform for quantum
simulations. Controlled generation of synthetic gauge
fields\cite{Lin2009A, Lin2009B, Lin2011} has provided impetus to the
realization of novel phases in cold atomic systems. The recent
generation of synthetic non-Abelian gauge fields in $^{87}$Rb
atoms\cite{Lin2011} is a key step forward in this regard. While a
uniform Abelian gauge field is merely equivalent to a galilean
transformation, even a uniform non-Abelian gauge field
nurtures interesting physics.\cite{Ho2010,Wang2010,Lin2011}

The clue that a uniform non-Abelian gauge field crucially influences
the physics of interacting fermions came from the study of bound
states of two spin-$\half$ fermions in its presence.\cite{Vyasanakere2011}
The remarkable result found for spin-$\half$ fermions in three spatial
dimensions interacting via a $s$-wave contact interaction in the
singlet channel is that high-symmetry non-Abelian gauge field configurations
(GFCs) induce a two-body bound state for {\em any} scattering length
however small and negative. The physics behind this unusual role of
the non-Abelian gauge field that produces a generalized Rashba
spin-orbit interaction, was explained by its effect on the infrared
density of the states of the noninteracting two-particle spectrum. The non-Abelian
gauge field drastically enhances the infrared density of states, and
this serves to ``amplify the attractive interactions''. A second most
remarkable feature demonstrated in ref.~\myonlinecite{Vyasanakere2011} is
that wave function of the bound state that emerges has a triplet
content and associated spin-nematic structure similar to those found
in liquid $^3$He.

The above study\cite{Vyasanakere2011} motivated the study of
interacting fermions at a {\em finite} density in the presence of a
non-Abelian gauge field.\cite{Vyasanakere2011b} At a finite density
$\rho$ ($\sim k_F^3$, $k_F$ is the Fermi momentum), the physics of
interacting fermions in a synthetic non-Abelian gauge field is
determined by two dimensionless scales. The first scale is associated
with the size of the interactions $-1/k_F\as$ where $\as$ is the
$s$-wave scattering length, and the second one, $\frac{\lambda}{k_F}$,
is determined by the non-Abelian gauge coupling strength $\lambda$. For small
negative scattering lengths ($-1/k_F\as \gg 1)$, the ground state in
the absence of the gauge field is a BCS superfluid state with large
overlapping pairs. The key result first demonstrated in
ref.~\myonlinecite{Vyasanakere2011b} is that at a {\em fixed scattering
  length}, even if small and negative, the non-Abelian gauge field
induces a crossover of the ground state from the just discussed BCS
superfluid state to a new type of BEC state.  The BEC state that
emerges is a condensate of a collection of bosons which are tightly
bound pairs of fermions. Remarkably, at large gauge couplings
$\lambda/k_F \gg 1$, the nature of the bosons that make up the
condensate is determined {\em solely by the gauge field} and is not
influenced by the scattering length (so long as it is non-zero), or by
the density of particles. In other words, the BEC state that is
attained in the $\lambda/k_F \gg 1$ regime at a fixed scattering
length does not depend on the value of the scattering length, i.~e.,
the BEC is a condensate of a novel bosonic paired state of fermions
determined by the non-Abelian gauge field.  These bosons were called
as ``rashbons'' since their properties are determined solely by the
generalized Rashba spin-orbit coupling produced by the gauge field. As
shown in ref.~\myonlinecite{Vyasanakere2011b}, rashbon is the bound
state of two fermions at infinite scattering length (resonance) in the
presence the non-Abelian gauge field. 
The crossover from the BCS state to the ``rashbon BEC'' state (RBEC)
induced by the gauge field at a fixed scattering length is to be
contrasted with the traditional BCS-BEC
crossover\cite{Eagles1969,Leggett1980,Randeria1995,Leggett2006} by
tuning the scattering
length\cite{Regal2004,Ketterle2008,Giorgini2008}, but with no gauge
field. Gong et al.\cite{Gong2011} have investigated the crossover
including the effects of a Zeeman field along with a non-Abelian gauge
field. Certain properties of rashbons in the EO gauge field (explained
later) have been investigated in references \myonlinecite{Hu2011} and
\myonlinecite{Yu2011}.

\begin{figure}
\includegraphics[width=\myfigwidth]{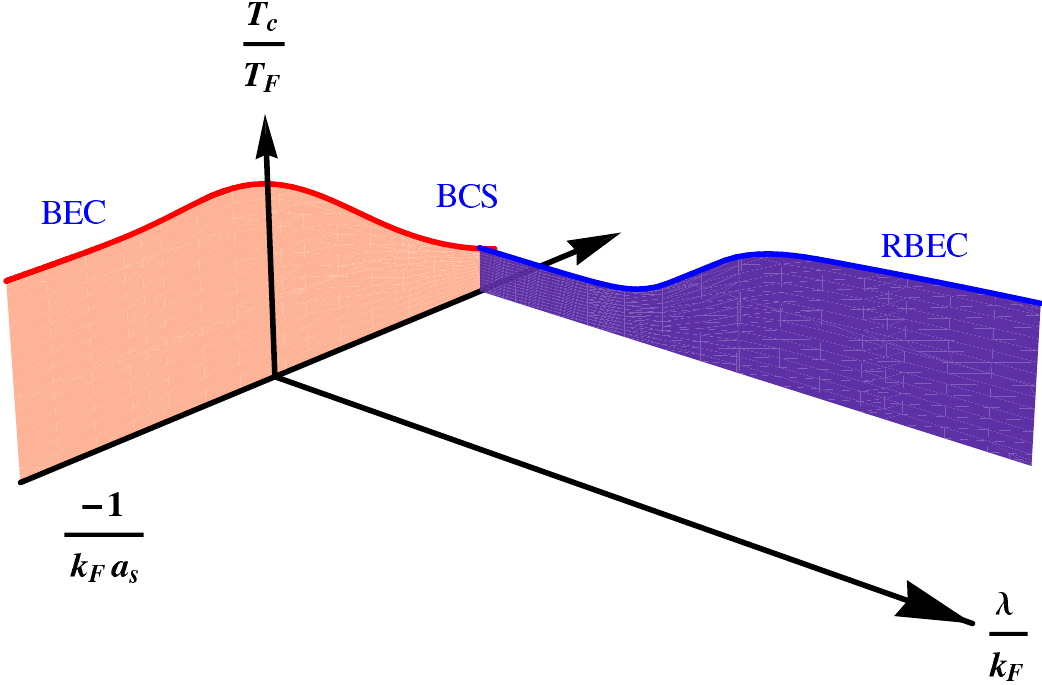}
\caption{BCS-RBEC crossover induced by a non-Abelian gauge field. Here, $\as$ is the $s$-wave scattering length, $k_F$ is the Fermi momentum determined by the density, $T_F$ is the Fermi temperature ($=k_F^2/2$), $T$ is the temperature $\lambda$ is the strength of the gauge coupling. The solid red line is the transition temperature of the superfluid phase (shaded in light red) obtained in ref.~\onlinecite{SaDeMelo1993} using the Nozi\'eres-Schmitt-Rink theory\cite{Nozieres1985}. The solid blue curve is based on the estimate presented in this work. The figure reveals the qualitative features of the full ``phase diagram'' in the $T-\as-\lambda$ space. (Figure courtesy: Sudeep Kumar Ghosh) }
\label{fig:Sudeep}
\end{figure}

It was shown in ref.~\myonlinecite{Vyasanakere2011b} that the Fermi surface
of the non-interacting system (with $\as =0)$ in presence of the
non-Abelian gauge field undergoes a change in topology at a critical
gauge coupling strength $\lambda_T$ (of order $k_F$). For weak
attractions $(-1/k_F\as \gg 1)$, the regime of the gauge coupling
strengths where the crossover from  the BCS state to the RBEC state takes
place coincides with the regime where the bare Fermi surface undergoes
the topology change. The properties of the superfluid state (such as
the transition temperature) for $\lambda \gtrsim \lambda_T$ was argued
to be primarily determined by the properties of the constituent
anisotropic rashbons (see Sect.~V of
ref.~\myonlinecite{Vyasanakere2011b}). It is, therefore, necessary and
fruitful to undertake a detailed study of the properties of rashbons
and their dispersion, and this is the aim of this paper.

In this paper, we study the properties of rashbons and their
dependence on the nature of the non-Abelian gauge field, i.e., we
obtain properties of rashbons for the most interesting gauge field
configurations.  This study entails a study of the anisotropic rashbon
dispersion, i.~e., determination of its energy as a function of its
momentum by the study of the two-body problem in a non-Abelian gauge
field with a resonant scattering length ($1/\lambda \as= 0$). In
addition to the determination of the properties of rashbons, we report
here a new qualitative result. It is shown that {\em when the momentum
  of a rashbon exceeds a critical value which is of the order of the
  gauge coupling strength, it ceases to exist}. Stated otherwise, when
the center of mass momentum of the two fermions that make up the bound
pair exceeds a value of order of the gauge coupling strength, the bound
state disappears. To uncover the physics
behind this result, the two-fermion problem in a gauge field is
investigated in detail for a range scattering of lengths and centre of
mass momenta. The study reveals a hitherto unknown feature of the
non-Abelian gauge fields: while the non-Abelian gauge field acts as
attractive interaction amplifiers for fermions with centre of mass
momenta $q$ much smaller than the gauge field strength ($q \ll
\lambda$), the gauge field {\em suppresses the formation of bound
  states of fermions with large centre of mass momenta ($q \gtrsim
  \lambda$)}. In fact, it is demonstrated here that when $q \gtrsim
\lambda$, a {\em positive} scattering length (very strong attraction)
is necessary to induce a bound state of the two fermions, quite
contrary to $q \ll \lambda$ where a bound state exists (essentially)
for any scattering length.

The results we report here  have two significant outcomes. (1) A full
qualitative picture of the BCS-BEC crossover scenario in the presence
of a non-Abelian gauge field is obtained (see \Fig{fig:Sudeep}) based
on the results reported here. Most notably, it is shown that the
transition temperatures of a system of fermions with a very weak
attraction can be enhanced to the order of the Fermi temperature 
(determined by the density) by the application of a non-Abelian gauge
field. (2) Our two body results at large centre of mass momenta
suggest that the normal state of the fermion system in non-Abelian
gauge field will be a ``dynamic mixture'' of rashbons and interacting
helical fermions. These could therefore show many novel features such
as pronounced pseudogap characteristics (see ref.~\onlinecite{Mueller2011} and references therein).

The next section, \ref{sec:Prelims}, contains the preliminaries
which includes the formulation of the
problem. Sec.~\ref{sec:Rashbons} contains a report on the
properties of rashbons, and this is followed by
sec.~\ref{sec:Specific} which discusses the bound state of two
fermions for arbitrary centre of mass momentum and scattering length for
specific high symmetry gauge fields. The importance of the results
obtained here is discussed in sec.~\ref{sec:Significance}, and the
paper is concluded with a summary in sec.~\ref{sec:Summary}.

\section{Preliminaries}
\mylabel{sec:Prelims}

 The Hamiltonian of the fermions moving in a uniform non-Abelian gauge field that leads to a generalized Rashba spin-orbit interaction is\footnote{A more detailed classification of the non-Abelian gauge fields can be found in Ref. \myonlinecite{Vyasanakere2011,Vyasanakere2011b}.} 
\beq\mylabel{eqn:Rashbha}
{\cal H}_R = \int \D{^3 \br} \, \Psi^\dagger(\br) \left( \frac{\bp^2}{2} \bOne - \bp_\lambda \cdot \btau \right) \Psi(\br),
\eeq
where $\Psi(\br) = \{\psi_{\sigma}(\br)\}, \sigma= \uparrow, \downarrow$ are fermion operators, $\bp$ is the momentum, $\bOne$ is the SU(2) identity, $\tau^\mu$ ($\mu=x,y,z$) are Pauli matrices, $\bp_\lambda = \sum_i p_i \lambda_i \be_i$, $\be_i$'s are the unit vectors in the $i$-th direction, $i=x,y,z$. The vector $\blam =\lambda \hat{\blam} = \sum_i \lambda_i \be_i$ describes a gauge-field configuration (GFC) space; we refer $\lambda = |\blam|$ as the gauge-coupling strength. Throughout, we have set the mass of the fermions ($m_F$), Planck constant ($\hbar$) and Boltzmann constant ($k_B$) to unity.

In this paper we specialize to $\blam = (\lambda_l,\lambda_l,\lambda_r)$ as this contains all the experimentally interesting high-symmetry GFCs. Moreover, it is shown in Ref. \myonlinecite{Vyasanakere2011,Vyasanakere2011b}, that this set of gauge fields captures all the qualitative physics of the full GFC space. Specific high symmetry GFCs are obtained for particular values of $\lambda_r$ and $\lambda_l$:  $\lambda_r=0$ corresponds to extreme oblate (EO) GFC; $\lambda_r=\lambda_l$ corresponds to spherical (S) GFC,  and $\lambda_l=0$ corresponds to extreme prolate (EP) GFC.

The interaction between the fermions is described by a contact attraction in the singlet channel 
\beq\mylabel{eqn:Interaction}
{\cal H}_{\upsilon} = \upsilon \int \D{^3 \br}\, \psi^\dagger_\uparrow(\br) \psi^\dagger_{\downarrow}(\br) \psi_{\downarrow}(\br) \psi_{\uparrow}(\br).
\eeq
Ultraviolet regularization\cite{Braaten2008} of the theory described by ${\cal H} = {\cal H}_R + {\cal H}_\upsilon$ is achieved by exchanging the bare interaction $v$ for the scattering length $\as$ via $\frac{1}{\upsilon} + \Lambda = \frac{1}{4 \pi \as}$, where  $\Lambda$  is the ultraviolet momentum cutoff. Note that $\as$ is the $s$-wave scattering length in free vacuum, i.~e., when the gauge field is absent ($\lambda=0$).

The one-particle states of ${\cal H}_R$ are described by the quantum
numbers of momentum $\bk$ and helicity $\alpha$ (which takes on values
$\pm$) : $\ket{\bk \alpha} = \ket{\bk} \otimes \ket{\alpha
  \hat{\bk}_\lambda}$. The one-particle dispersion is
$\varepsilon_{\bk \alpha} = \frac{k^2}{2} - \alpha |\bk_\lambda|$,
where $\bk_\lambda$ is defined analogously with $\bp_\lambda$ and
$\ket{\alpha \hat{\bk}_\lambda}$ is the spin-coherent state in the
direction $\alpha \hat{\bk}_\lambda$. The two-particle states of
${\cal H}$ can be described using the basis states $\ket{\bq \bk
  \alpha \beta} = \ket{(\frac{\bq}{2} + \bk) \alpha} \otimes
\ket{(\frac{\bq}{2} - \bk) \beta}$ where, $\bq = \bk_1 + \bk_2$ is the
center of mass momentum and $\bk = (\bk_1 - \bk_2)/2$ is the relative
momentum of two particles with momenta $\bk_1$ and $\bk_2$. Note that
$\bq$ is a good quantum number for the full Hamiltonian (${\cal
  H}$). The non-interacting two-particle dispersion is $E^{free}_{\bq
  \bk \alpha \beta} = \varepsilon_{(\frac{\bq}{2} + \bk) \alpha} +
\varepsilon_{(\frac{\bq}{2} - \bk) \beta}$. In the presence of
interactions, bound states emerge as isolated poles of the $T$-matrix,
and are roots of the equation
\beq\mylabel{eqn:secular}
\frac{1}{4 \pi a_s} =  \frac{1}{V} \sum_{\bk \alpha \beta} \left( \frac{|A^{\bq}_{\alpha \beta}(\bk)|^2}{E(\bq)-E^{free}_{\bq \bk \alpha \beta}} + \frac{1}{4 k^2} \right)
\eeq
where, $A^{\bq}_{\alpha \beta}(\bk)$ is the singlet amplitude in $\ket{\bq \bk \alpha \beta}$, $V$ is the volume, $E(\bq)=E_{th}(\bq)-E_b(\bq)$ is the energy of the bound state. Here $E_{th}(\bq)$ is the scattering threshold and $E_b(\bq)$ is the binding energy, both of which depend on $\bq$ as indicated.

In the absence of the gauge field ($\lambda =0$), the bound state
exists only for $\as>0$ and $E_b(\bq) = -1/\as^2$ is independent of
$\bq$. The threshold is $E_{th}(\bq)=q^2/4$. Physically, this
corresponds to the fact that a critical attraction in necessary in
free vacuum ($\lambda=0$) for the formation of the two-body bound state. As shown in
ref.~\myonlinecite{Vyasanakere2011}, state of affairs change
drastically in the presence of a non-Abelian gauge field. For $\bq=0$,
the presence of the gauge field always reduces the critical attraction
to form the bound state and in particular, for special high symmetry
GFCs (e.g. $\blam = (\lambda_l,\lambda_l,\lambda_r)$ with $\lambda_r
\leq \lambda_l$) two body bound state forms for any scattering
length.\cite{Vyasanakere2011}

\section{Properties of Rashbons}
\mylabel{sec:Rashbons}

\begin{figure}
\centerline{{\large $~~~~~~~~~~$ {\sf EO} $~~~~~~~~~~~~~~~~~~~~~~~~~${\sf S} $~~~~~~~~~~~~~~~~~${\sf EP}}}
\includegraphics[width=\myfigwidth]{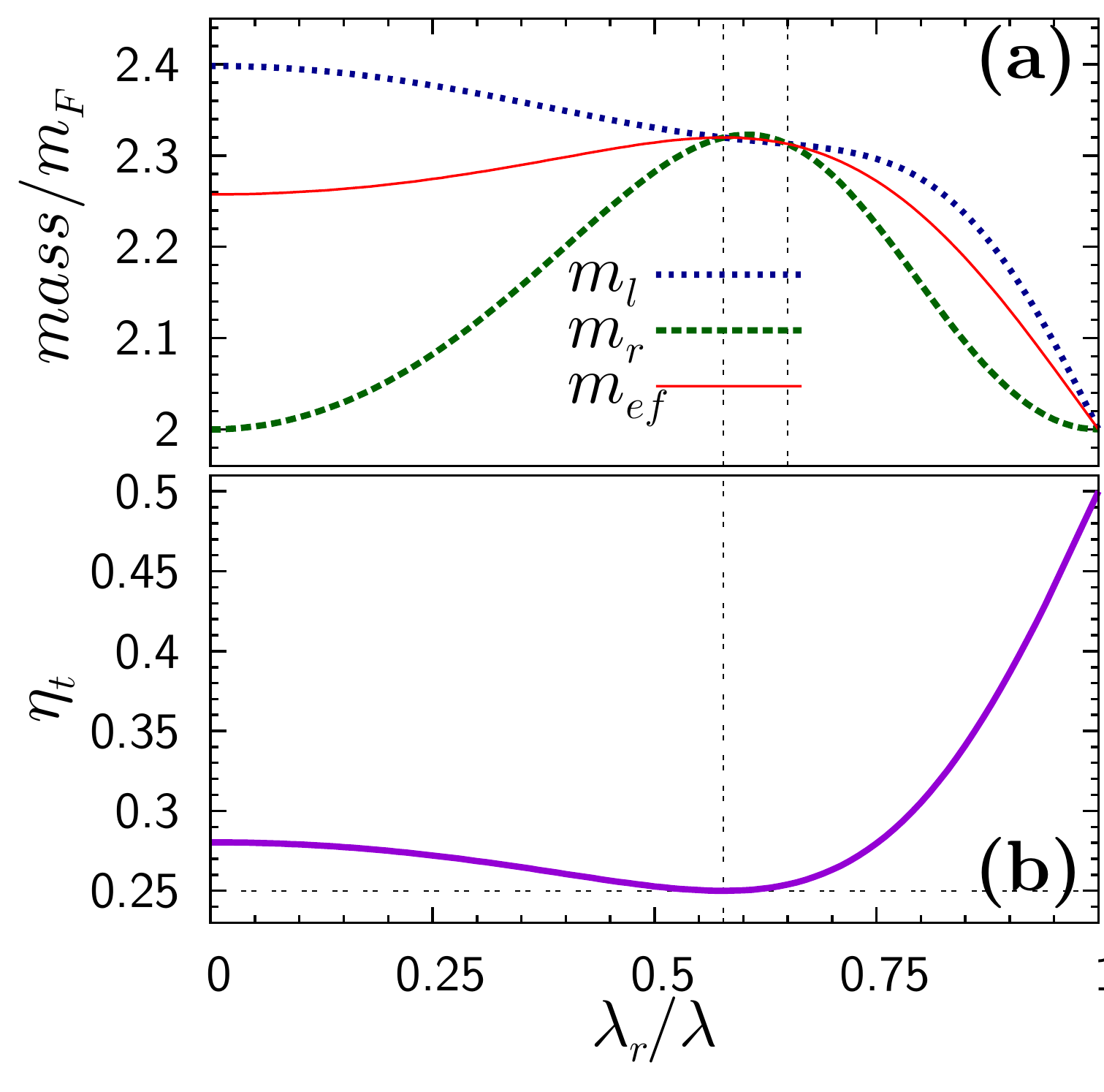}
\caption{(Color online) Rashbon properties for different GFCs. {\bf (a)} In-plane, perpendicular and effective masses. {\bf (b)} The triplet content of rashbons.}
\label{fig:Rashbon}
\end{figure}

The bound state that emerges in the presence of the gauge field when
the scattering length is set to the resonant value $1/\as = 0$ is the
rashbon. As argued above, the binding energy of the rashbon state for
all the GFCs considered here (except for the EP GFC) is positive. The
energy of the rashbon state $E_R(\bq=0)$ determines the chemical
potential of the RBEC. Other properties of the RBEC are determined by
the  rashbon dispersion $E_R(\bq)$, and in particular the
transition temperature will be determined by the mass of the rashbons.

The curvature of the rashbon dispersion $E_R(\bq)$ at $\bq=\bzero$ 
defines the effective low-energy inverse mass of rashbons. The dispersion is in general anisotropic and the inverse mass is, in general, a tensor.
However, due to their symmetry, for the GFCs considered in this
paper ($\blam$ of the form $(\lambda_l,\lambda_l,\lambda_r)$),
$E_R(\bq)=E_R(q_l,q_r)$, where $q_l$ is the component of $\bq$ on the
 $x-y$ plane, and $q_r$ is the component
along $\be_z$. Thus, the inverse mass tensor is completely specified by
its principal elements - in-plane inverse mass ($m_l^{-1}$) and the ``perpendicular'' inverse mass, $m_r^{-1}$. 
\beq\label{eqn:RashbonMass}
m_l^{-1} = \left. \frac{\dou^2 E_R(q_l, q_r)}{\dou q_l^2}\right|_{\bq = 0},\;\;\;\;m_r^{-1} = \left. \frac{\dou^2 E_R(q_l, q_r)}{\dou q_r^2}\right|_{\bq = 0}
\eeq
An effective mass $m_{ef}$ defined as
\beq
m_{ef} = \sqrt[3]{(m_r m_l^2)}
\eeq
is useful in the discussions that follow.

In addition to the anisotropy in their orbital motion, rashbons are
intrinsically anisotropic particles. Their pair-wave function has both
a singlet and triplet component; the weight of the pair wave function
in the triplet sector $\eta_t$ is the triplet content. The triplet
component is time reversal symmetric, but does not have the spin
rotational symmetry -- it is therefore a spin nematic. Keeping this interesting aspect in
mind, we shall also investigate and report the triplet content of
rashbons, and its dependence on the gauge field.

Before presenting the results we make a general observation. The
threshold energy ($E_{th}$) becomes increasingly flat as a function of
$\bq$ in the small $\bq/\lambda$ regime as one approaches spherical
gauge field in the GFC space. In fact, for the spherical GFC, it is
exactly constant in the small $\bq/\lambda$ regime (see
\Fig{fig:Eb_smallq}). The mass is therefore determined entirely by the
variation of the binding energy with $\bq$ (this may be contrasted in
free vacuum case discussed before). It is reasonable therefore to
expect that the effective mass of rashbons is always greater than
twice the bare fermion mass and for it to be the largest for the
spherical GFC.

\Fig{fig:Rashbon} {\bf (a)} shows the in-plane, perpendicular, and effective masses for different GFCs.  Rashbons emerging
from spherical GFCs have the highest $m_{ef}$ and that from EP GFCs
have the least. It is interesting to note that apart from the
spherical GFC, there is yet another GFC ($\lambda_r \approx 0.65
\lambda$ - see \fig{fig:Rashbon} {\bf (a)}) where the low energy
dispersion is isotropic, i.~e., rashbon has a scalar mass. The
triplet content is shown in \fig{fig:Rashbon} {\bf (b)} for different
GFCs. $\eta_t$ is minimum (1/4) for spherical GFC and maximum (1/2) for EP
GFC.


A detailed study of rashbon dispersion as a function of its momentum
$\bq$ (centre of mass momentum of the fermions that make up the
rashbon) revealed a hitherto unreported and rather unexpected
feature. The full rashbon dispersion as a function of $\bq$ for the
spherical (S) GFC is shown in \Fig{fig:Eb_smallq}. The rashbon energy
increases with increasing $q$ and eventually for $q/\lambda \gtrsim
1.3$, there is no two body bound state! This curious result motivated
us to perform a more detailed investigation of the dispersion of the
bound fermions (bosons) at arbitrary scattering lengths (away from 
resonance which corresponds to rashbons), in order to uncover the
physics behind this phenomenon. This study, conducted for specific high symmetry GFCs, is presented in the next section.

\section{Dispersion of bosons at arbitrary scattering lengths for specific GFCs}
\mylabel{sec:Specific}

In this section we investigate the dispersion of the bosonic bound state of two fermions at arbitrary scattering length. Results of the boson dispersion obtained by solving \eqn{eqn:secular} will be presented for the S and EO GFCs.

\subsection{Spherical GFC}

\begin{figure*}
\centerline{
\includegraphics[width=\myfigwidth]{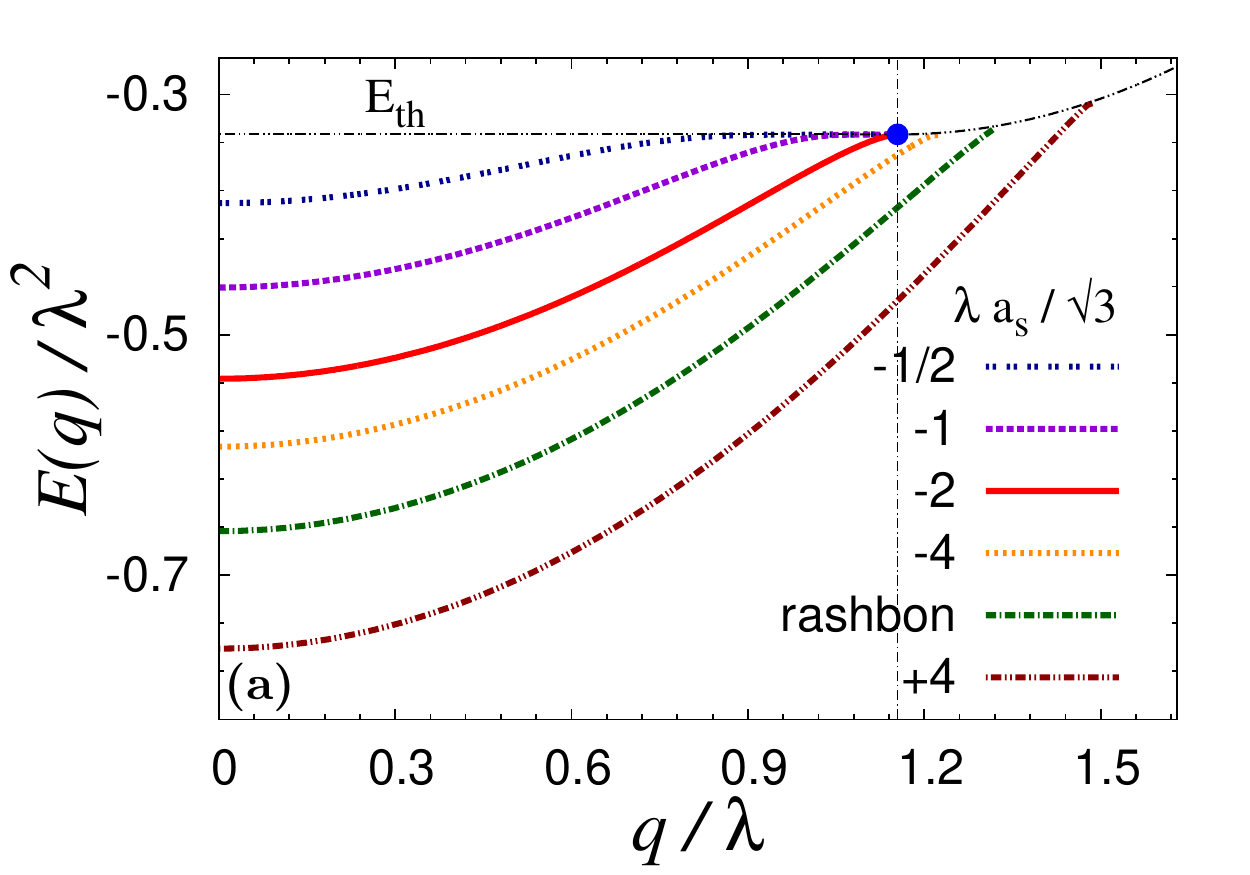}
\includegraphics[width=\myfigwidth]{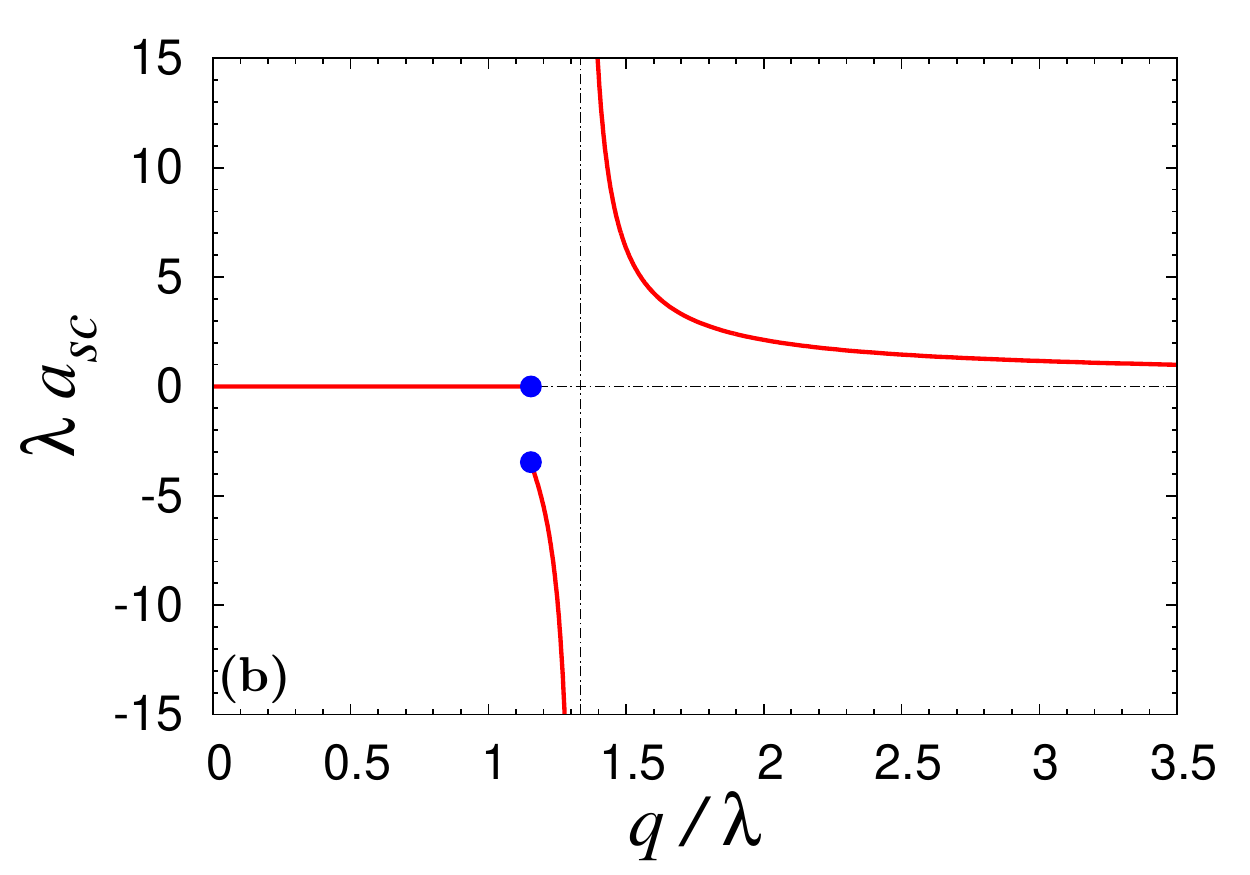}}
\caption{(Color online) {\bf (a)} The boson dispersion for various scattering lengths in spherical GFC.  Note that for any given scattering length, the bound state disappears after some critical momentum. {\bf (b)} Critical scattering length ($\asc$) as a function of momentum. $\asc$ goes as $1/\sqrt{q}$ in the large $q/\lambda$ limit.}
\label{fig:Eb_smallq}
\end{figure*}

Spherical (S) GFC corresponds to $\lambda_r=\lambda_l$ and hence
produces an isotropic boson dispersion as discussed before. The boson
dispersion depends only on $q=|\bq|$. Solving \eqn{eqn:secular}, the
boson dispersion obtained for various scattering lengths is as shown
in \fig{fig:Eb_smallq}(a). The key features of this spectrum are the
following.  For {\em any scattering length, however large and
  positive}, there exists a critical center of mass momentum $q_c$
such that when $q > q_c$ the bound state ceases to exist. 

This is best understood by fixing attention on a particular momentum
$q$. When the momentum is ``small'', there is a bound state for {\em any
  attraction}. This is in fact the case for all $q < q_o$, where $q_o
= 2 \frac{\lambda}{\sqrt{3}}$. For $q > q_o$, a critical attraction
described by a nonzero scattering length $\asc$ is necessary for the
formation of a bound state. For $q = q_o^+$, the critical scattering
length is $\asc = -\frac{2 \sqrt{3}}{\lambda}$. On increasing $q$, a 
stronger attractive interaction is required to produce a bound state
and when $q$ reaches $\sim \frac{4\lambda}{3}$, a resonant attraction is
necessary to produce a bound state. For $q \gtrsim \frac{4\lambda}{3}$,
a very strong attractive interaction described by a small positive
scattering length is necessary to produce a bound state. In fact, for
$q \gg \lambda$, the critical scattering length scales as $\asc \sim \sqrt{\frac{1}{\lambda q}}$. The dependence of $\asc$ on the centre of mass momentum is shown in \Fig{fig:Eb_smallq}(b).

How do we understand these results? Here the $\varepsilon_0-\gamma$
model introduced in Ref.~\onlinecite{Vyasanakere2011} comes to our
rescue. The model states that if the infrared density of states
$g_s(\varepsilon) \sim \varepsilon^\gamma$ for $0 \le \varepsilon \le
\varepsilon_0$, where $\varepsilon$ is the energy measured from the
scattering threshold, then the critical scattering length is given by
$ \sqrt{\varepsilon_0} \asc \propto \gamma \, \Theta(\gamma) / (2
\gamma -1) $, where $\Theta$ is the unit step function. Note that for
$\gamma < 0$, the critical scattering length vanishes.

It is evident that there is a drastic change in the infrared density of
states at $q=q_o$. In fact, this special momentum $q_o$ is
such that the threshold energy corresponds to that state where the
relative momentum $\bk$ between the pair of fermions
vanishes. Clearly, for $q < q_o$, there are many degenerate $\bk$ states that
produce a nonzero density of states at the threshold. In fact, when $q
=0$, the density of states diverges as $1/\sqrt{\varepsilon}$, i.e.,
$\gamma = -1/2$. For $q < q_o$, there is still a finite density of
states at the threshold with an effective $\gamma< 0$. Thus the
critical scattering length, as given by the $\epsilon_0-\gamma$ model,
vanishes. Let us turn our attention to what happens for $q \gg
\lambda$. From \eqn{eqn:secular} it is evident that the density of
states $g_s(\varepsilon)$ has the contributions from the $++$, $--$,
$+-$ and $-+$ channels. It can be shown that in the regime $q \gg
\lambda$, of the $++$ channel has a density of states that has $\varepsilon^{3/2}$ behaviour. The
$+-$ and $-+$ channels have a higher threshold which is $\lambda q$
larger than the threshold of the $++$ channel; the density of
states of the $+-$/$-+$ channels goes as $\sqrt{\varepsilon}$ form
this higher threshold. These arguments provide an estimate of
$\varepsilon_0 \approx q \lambda$. The result on the critical
scattering length is then $\asc \sim \frac{1}{\sqrt{q}}$, precisely as
obtained from the full numerical solution shown in
\Fig{fig:Eb_smallq}(b).

As a by product of the analysis of the boson dispersion, we were able to obtain an analytical expression for the mass of bosons (which is isotropic in this case)
\beq
\frac{m_B}{m_F} = \frac{6 E(\bzero)^{3/2}}{7 E(\bzero)^{3/2} + 2 \sqrt{E(\bzero)} \lambda ^2-4
   \left( E(\bzero) + \frac{\lambda ^2}{3} \right)^{3/2}}, \quad
\label{eqn:mass_anlt}
\eeq
where,
\beq
E(\bzero) = -\frac{\lambda^2}{3}-\frac{1}{4} \left(\frac{1}{a_s} + \sqrt{\frac{1}{a_s^2} + \frac{4\lambda^2}{3}}\right)^2
\eeq
At a given $\lambda$, as expected, mass for a small positive scattering length $\as > 0$  is twice the fermion mass. Mass at resonance is the  rashbon mass which is equal to $=\frac{3}{7} (4+\sqrt{2}) m_F \approx 2.32 m_F$. Interestingly, the value of $m_B/m_F$ for a small negative scattering length limit is (integer) $6$.

\subsection{Extreme oblate GFC}

\begin{figure}
\includegraphics[width=\myfigwidth]{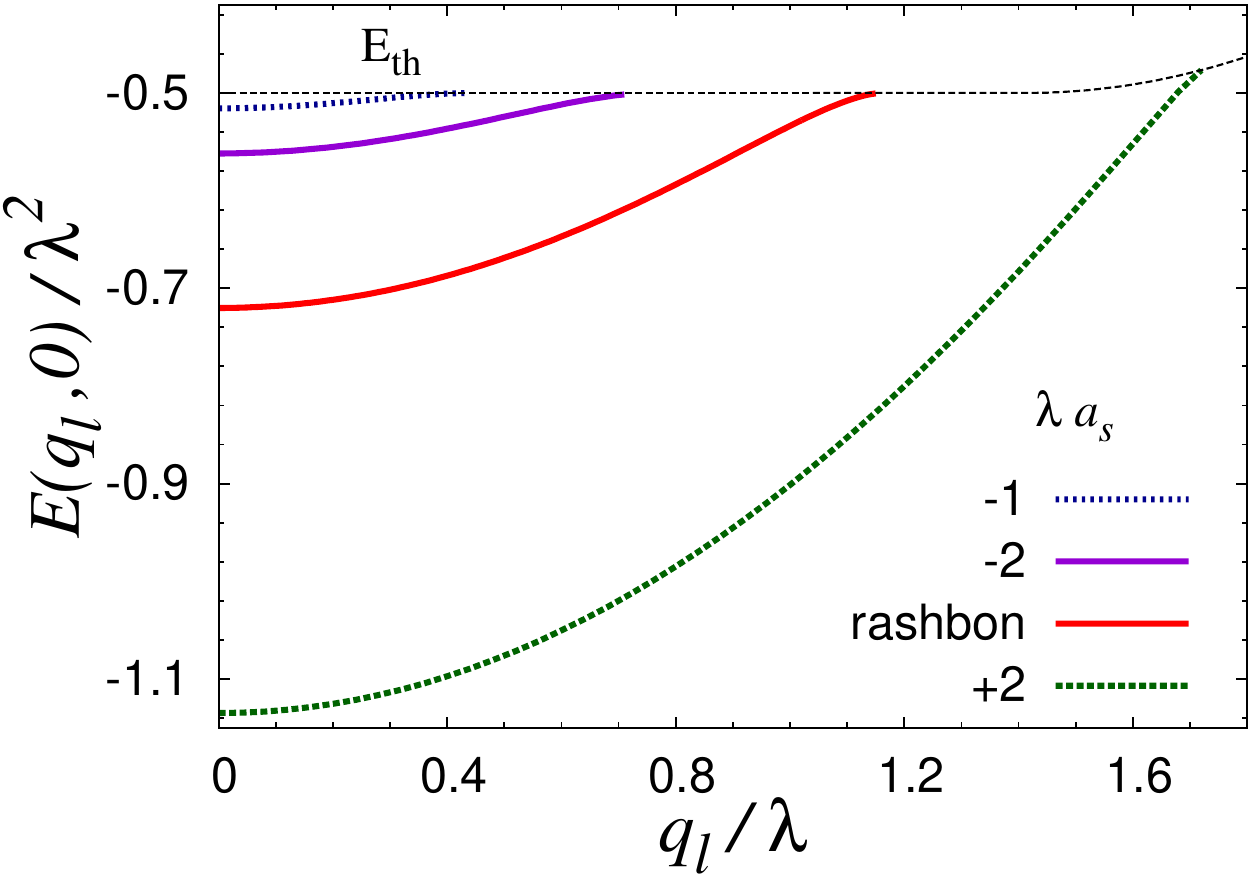}
\caption{(Color online) The boson dispersion for various scattering lengths in extreme oblate GFC.  Just as in S GFC (see \Fig{fig:Eb_smallq}), for any given scattering length, the bound state disappears after some critical $q_l$.}
\label{fig:Dispersion_EO}
\end{figure}

\begin{figure}
\includegraphics[width=\myfigwidth]{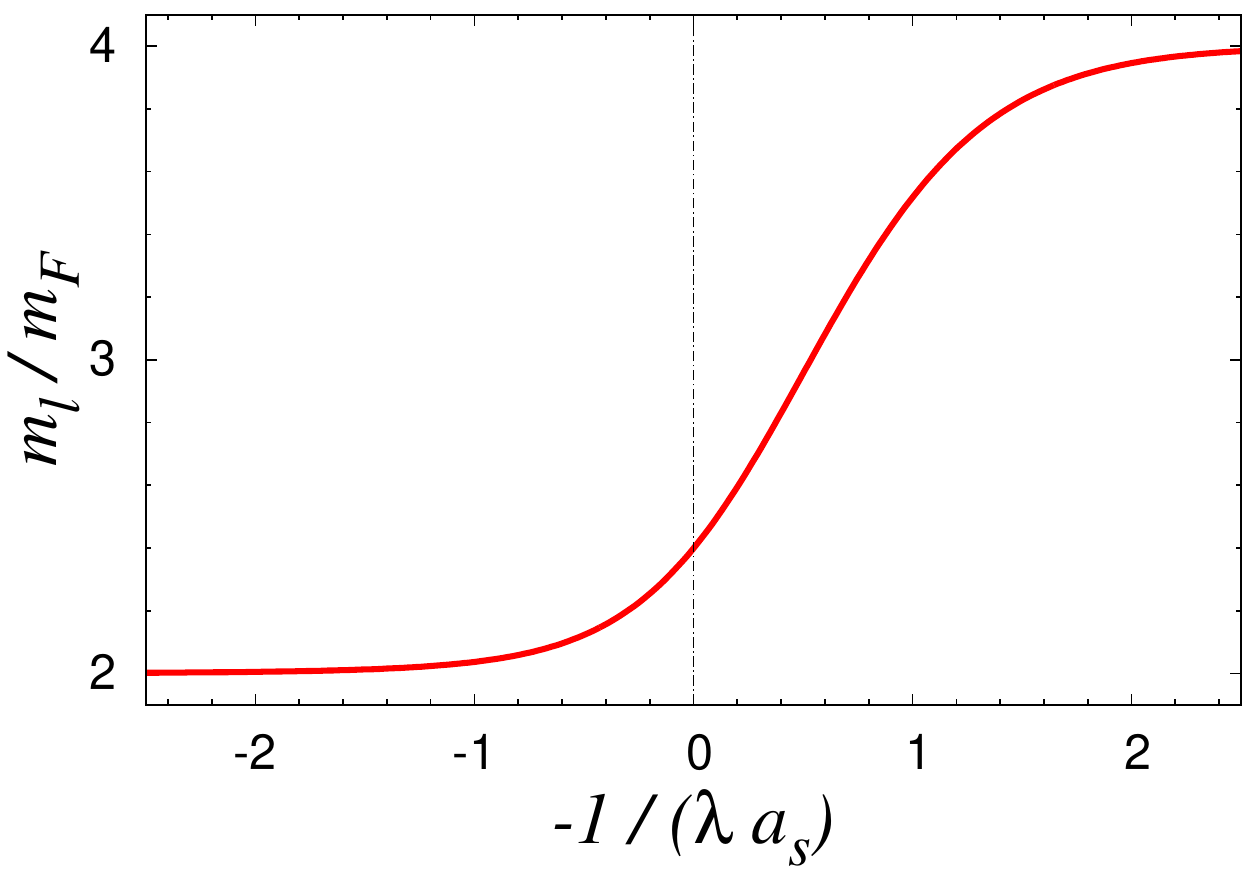}
\caption{(Color online) In-plane mass of tightly bound fermion pairs in the rashbon BEC side in presence of an extreme oblate GFC. }
\label{fig:Tc_and_mB_EO}
\end{figure}

Extreme oblate (EO) GFC corresponds to $\lambda_r=0$ with $\lambda_l = \frac{\lambda}{\sqrt{2}}$. It can be easily shown that for this GFC, $E(q_l,q_r)=E(q_l,0) + \frac{q_r^2}{2}$. Thus, the two-body dispersion as a function of $q_l$  provide all the nontrivial features of the two-body problem arising from this gauge field. 

\Fig{fig:Dispersion_EO} shows the boson dispersion for various scattering lengths. Remarkably, we find that the dispersion has very similar features as found for the spherical GFC, i.e., for any given scattering length there is a $q_c$ such that for $q > q_c$, the two-body bound state ceases to exist. Clearly, this is a generic feature of the boson (bound fermion-pair) dispersion in a gauge field.

For this GFC, $m_r$ is just twice the fermion mass. The in-plane mass ($m_l$) extracted from the two-body dispersion is shown in \fig{fig:Tc_and_mB_EO}. $m_l$ for small positive scattering length  is again twice the fermion mass. The resonance value which  corresponds to rashbon is $m_l\simeq 2.4 m_F$. This result agrees with refs.~\myonlinecite{Yu2011,Hu2011}. It is again interesting to note that value of $m_l/m_F$ in the deep BCS limit is (integer) 4.

\subsection{Discussion}

The analysis of the dispersion of the boson (bound-state of two
fermions obtained in a gauge field)  reveals that the boson
ceases to exist when the momentum of the boson exceeds a critical
value. For the case of rashbons (bosons obtained at resonance
scatteirng length), the critical momentum is of the order of the
strength of the gauge field.

The analysis presented here shows that this is again because of the
influence of the gauge field in altering of the infrared density of
states. When the momentum is smaller than the magnitude of the gauge
coupling, the gauge field works to {\em enhance} the infrared density
of states. On the other hand, for large momenta, the gauge field has
the opposite effect, i.~e., it {\em depletes } the infrared density of
states.


\section{Significance of the Results}
\mylabel{sec:Significance}

The above results allow us to infer many key aspects of the physics of
interacting fermions in the presence of a non-Abelian gauge field.

First,  these results allow us to estimate the transition
temperature. For large gauge couplings, the transition temperature as
noted above will be determined by the mass of the rashbons. We have
argued (and demonstrated) that the mass of the rashbons is always
greater than twice the fermion mass. Thus the transition temperature
of RBEC will always be less than that of the usual BEC of bound pairs
of fermions obtained in the absence of the gauge field by tuning the
scattering length to small positive values.

However, there is something remarkable that a synthetic non Abelian
gauge field can achieve. Consider a system with a weak attraction
(small negative scattering length). In the absence of the gauge field,
the transition temperature in the BCS superfluid state is
exponentially small in the scattering length. Interestingly, the
transition temperature can be brought to the order of Fermi
temperature by increasing the magnitude of the gauge field strength
(keeping the weak attraction, small negative scattering length,
fixed). 
\begin{figure}
\includegraphics[width=\myfigwidth]{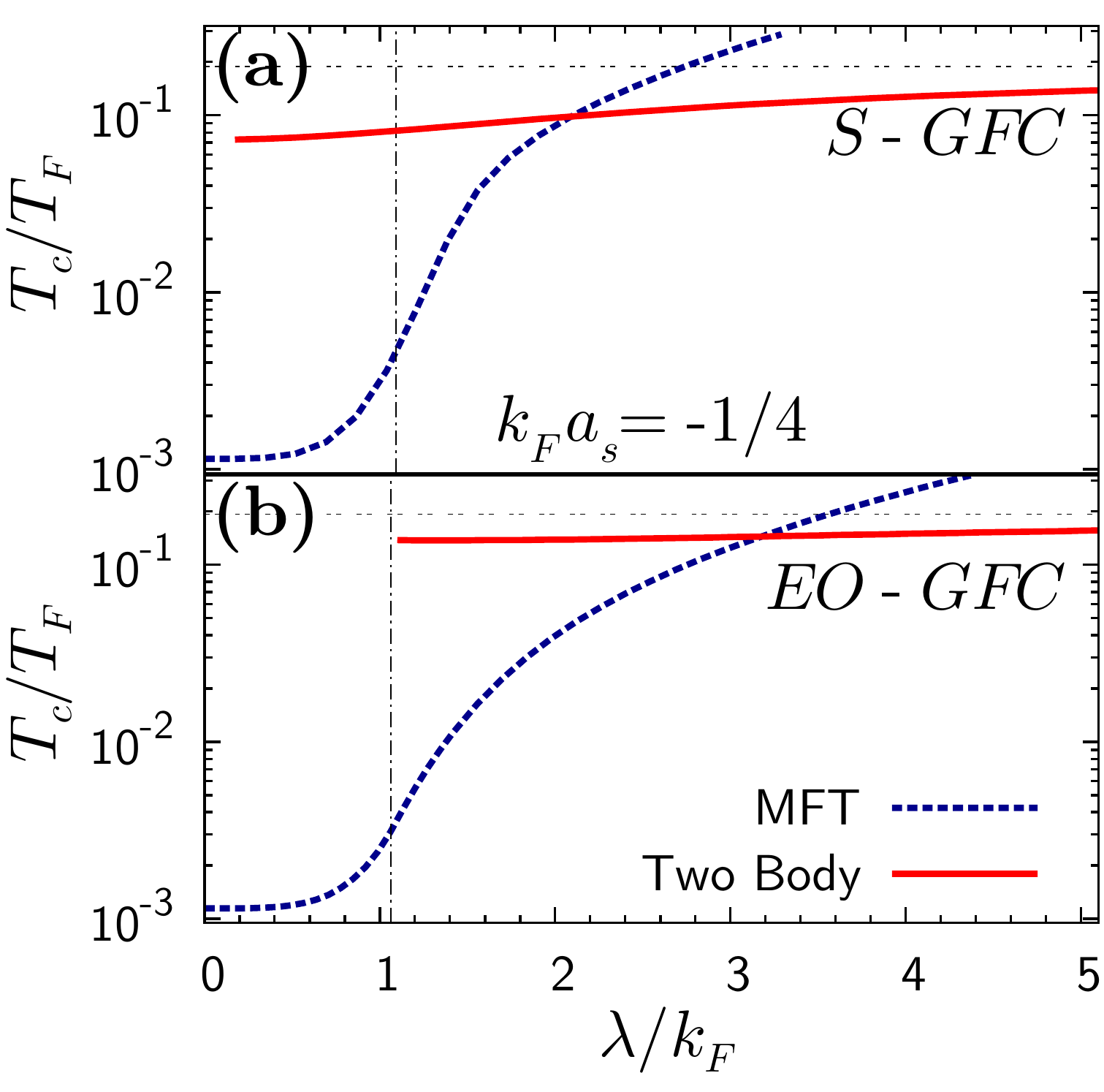}
\caption{(Color online) Estimate of transition temperature in spherical and extreme oblate GFCs as a function of the gauge coupling strength which takes the regular BCS state to a rashbon BEC. $T_c$ in the small $\lambda/\kf$ limit is obtained from mean field theory (analytical approximation is shown in the text). $T_c$ in the large $\lambda/\kf$ limit is obtained from the condensation temperature of the tightly bound pairs of fermions (analytical form for the S GFC can be obtained from \eqn{eqn:mass_anlt} and \eqn{eqn:T_BEC}). Horizontal dashed line corresponds to rashbon $T_c$. Vertical line indicates the gauge coupling corresponding to the Fermi surface topology transition\cite{Vyasanakere2011b}.}
\label{fig:Tc_m1by4_S_and_EO}
\end{figure}

 While $T_c$ in the BCS regime is determined by the pairing
amplitude ($\Delta$), in the BEC regime it is determined by the
condensation temperature of the emergent rashbons.\cite{Nozieres1985}

The mean field estimate of the former (i.e. for small $\kf |\as|$, $\as<0$ and small $\lambda / \kf$) is obtained by simultaneously solving $-1/(4 \pi \as) = \frac{1}{2V} \sum_{\bk \alpha} \left( \frac{\tanh{\frac{\xi_{\bk \alpha}}{2 T_c}}}{2\xi_{\bk \alpha}} - \frac{1}{k^2} \right) $ and the number equation, $\rho = \frac{1}{V} \sum_{\bk \alpha} 1/\left( \exp{\left(\frac{\xi_{\bk \alpha}}{T_c}\right)} + 1 \right),$ where $\xi_{\bk \alpha} = \varepsilon_{\bk \alpha} - \mu$. In this limit, the chemical potential at $T_c$ is almost equal to that of the noninteracting one at zero temperature. i.e., $ \mu(T_c,\as,\lambda) \approx \mu(0,0^-,\lambda) $, and $ \Delta_{(T=0)}/ T_c \approx \pi / e^{\gamma} $ where \cite{Vyasanakere2011b} $\Delta_{(T=0)}$ is the pairing amplitude at zero temperature and $\gamma$ is Euler's constant ($\approx$ 0.577).

The $T_c$ on the RBEC side can be extracted from the effective mass ($m_{ef}$) as condensation temperature of the bosonic pairs :
\bea
\frac{T_{c}}{T_{F}} &=& \left( \frac{16}{9 \pi (\zeta(3/2))^2} \right)^{1/3} \frac{1}{m_{ef}},\; \; \; \; \text{(rashbon BEC)} \; \;
\label{eqn:T_BEC}
\eea
where we recall that $m_{ef}=(m_r m_l^2)^{1/3}$. Using the information of mass given earlier (\eqn{eqn:mass_anlt} for S GFC and \fig{fig:Tc_and_mB_EO} for EO GFC) one can obtain $T_c$ in this regime as a function of $\lambda \as$ in S and EO GFCs. In particular, rashbon $T_c$ in S case is $\approx 0.188 T_F$ and in EO case it is $\approx 0.193 T_F$. The rashbon $T_c$ can be obtained for various GFCs, using $m_{ef}$ shown in \fig{fig:Rashbon} (\textbf{a}).  Since, among all GFCs, the rashbon mass corresponding to S GFC is the largest, it also corresponds to condensate with the smallest $T_c$.

The results obtained in both BCS and RBEC limits for $\kf \as =
-1/4$ in S and EO GFCs are shown in \fig{fig:Tc_m1by4_S_and_EO}. We
can see, as advertised, that $T_c$ has increased by two orders of
magnitude with increasing gauge coupling strength $\lambda$. These
considerations also allow us to infer an overall qualitative ``phase
diagram'' in the $T-\as-\lambda$ space as shown in \Fig{fig:Sudeep}.

What is the  nature of the system above $T_c$? There is a regime
in the parameter space shown in \Fig{fig:Sudeep}, where the normal
state can be quite interesting. Consider for example $\lambda \approx
1.5 k_F$. The ground state will be ``very bosonic'' i.~e., a
condensate of rashbons in the zero momentum state. On heating the
system above the transition temperature $\lesssim T_F$, the system
becomes normal. Rashbons are excited to higher momenta states, and
eventually break up into the constituent fermions since there is no
bound state at higher momenta. There, should, therefore be a
temperature range where the sytem is a dynamical mixture of
uncondensed rashbons and high energy helical fermions -- a state that
should show many novel features such as, among other things, a
pseudogap.

\section{Summary}
\mylabel{sec:Summary}

The new results of this paper are:

\begin{enumerate}
\item A systematic enumeration of the properties of rashbons, including closed form anlytical formulae,  for various gauge field configurations.
\item A detailed study of the rashbon (boson) dispersion, which results in a new qualitative observation. Although a zero centre of mass momentum bound state exists for any scattering length for many GFCs, the bound state vanishes when the centre of mass momentum exceeds a critical value. Thus, although the gauge field acts to promote bound state formation for small momenta, it acts oppositely, i.~e., inhibits bound state formation for large momenta. We provide a detailed explanation of the physics behind the phonemon.
\end{enumerate}

These results allow us to make two important inferences.
\begin{enumerate}
\item For a fixed weak attractive interaction, the exponentially small transition temperature of a BCS superfluid can be enhanced by orders of magnitude to the order the Fermi temperature of the system by increasing the magnitude of the gauge coupling.
\item There is a regime of  $T-\as-\lambda$ parameter space where the normal phase of the system will have novel features.
\end{enumerate}

We hope that these results will stimulate further experimental and theoretical studies on this topic.

\paragraph*{Acknowledgements}

J.V. acknowledges support from CSIR, India via a JRF grant. V.B.S. is
grateful to DST, India (Ramanujan grant) and DAE, India (SRC grant)
for generous support. We are grateful to Sudeep Kumar Ghosh for
providing us with \Fig{fig:Sudeep}.

\bibliography{refRashbon_nagf}

\end{document}